# Digitalization of COVID-19 pandemic management and cyber risk from connected systems

Petar Radanliev, David De Roure, Max Van Kleek

What makes cyber risks arising from connected systems challenging during the management of a pandemic? Assuming that a variety of cyber-physical systems are already operational-collecting, analyzing, and acting on data autonomously-what risks might arise in their application to pandemic management? We already have these systems operational, collecting, and analyzing data autonomously, so how would a pandemic monitoring app be different or riskier?

Read More

Perhaps unsurprisingly, the answer to these questions depends on specific aspects of the design and deployment of the connected systems in question. If established security design rules are followed, and focus is placed on what is required for pandemic management, this risks will be minimized. If, however, due to time and resource pressure such as from an unfolding pandemic, security is ignored for practicality and speed, new systems could be more easily compromised, leading to the potential for later system failures at crucial times. If such systems are designed and operated in the first wave of a global pandemic such as COVID-19, during the second and subsequent waves of system failure could lead to unnecessary loss of lives. We outline the security design principles that would minimize the potential risk scenarios.

Since the COVID-19 pandemic, the focus on stringent personal data protection for preserving individual privacy has been made more complicated by the need to support public health efforts that necessitate some degree of global surveillance assisted with new digital technologies. This brings into attention Internet of Things (IoT) technologies for their ability to operate autonomously to collect, analyse, and share data about the physical environment, which makes them essential elements in the digitalization of pandemic management. The IoT represents technologies that can use sensors for detection, gather and analyze information, create meaningful insights and act. The action usually represents a tailored product or service or improves the efficiency of operational processes. One difference between the IoT and the Internet is that the IoT can be completely automated and autonomous; in contrast, the Internet is fundamentally designed for application that connect people. With IoT technologies, the role of humans as actors in the network is arguably diminished. Automated and autonomous IoT technologies trigger questions on the potential cyber risk arising from the integration of artificial intelligence and machine learning in the IoT network. We review how artificial intelligence and machine learning can enable pandemic management, while ethically assessing the cyber risk from increased deregulation of data standards in IoT devices and networks.

**What is IoT risk?**

The emergence of the Internet of Things (IoT) in the late 1990s presented a new and fast-evolving technology that was characterized by low cost-high value per unit to a world that was unprepared to assess the associated risks. Since its emergence, IoT risks have been measured with traditional risk assessment methodologies and frameworks. One problem with such assessments during global pandemics is that IoT presents very different types of risks than the traditional Internet. Thus, an essential step to make sense of IoT risks is to understand IoT risk vectors [1]. One risk vector is that IoT, unlike the Internet, does not necessarily require human intervention [2]. IoT products can use connected sensors to collect and analyze data, then aggregate data or present data in various formats and trigger further actions based on data interpretation processes. The main question is, if interested parties acquire access to big data collected from individuals during a pandemic, what further actions could be triggered based on the aggregated new data, or even just from the interpretations of the new data?

A second risk vector is represented by the multiple and diverse connection points used to access any given IoT ecosystem, e.g. door locks connected to home security systems, or smart home hubs that collect personal and sensitive data via the devices they're connected to. Since the rapid spread of IoT devices is partially triggered by the low cost of materials, production, installation, automated data collection and analysis, one could argue that one of the main strengths of IoT is represented by the low cost and the relative ease in the deployment of this technology. However, securing IoT devices and assuring a level of maintenance on a par with other critical systems, such as typical of the telecom industry, for example, would require an increase in the initial cost for production and deployment and possibly a significant one in the operational costs. This increase in cost could diminish the main competitive edge of IoT devices.  Also, in complex IoT ecosystems, it can be difficult to find truly independent risks and to establish an actual correlation between pertinent risks [3]. Given the low cost of IoT devices, risk managing the complex coupled IoT systems has proven challenging even before the pandemic. Therefore, the focus must be on enabling data collection for pandemic risk management, while simultaneously limiting the aggregation of personal data that could be used for alternative interpretations.

**Value of IoT in pandemic management**

Despite the emerging IoT risks and the incomplete understanding of the impact created by these new risks, it is the value of IoT infrastructure in pandemic management and the promises for optimization of existing pandemic monitoring costs that drive governments to accept unknown risks and technological challenges. On a set of virtually no global IoT risk standards and policies, we anticipate further fragmentation of the IoT ecosystems, in which risk assessment models are likely to be volatile, vendor dependent, and less transparent [4]. Some of the advantages of the foreseen increase in heterogeneity consist of opportunities through competitive innovation. However, innovation in the IoT space implies important variations and unknowns concerning critical aspects such as security, adoption, and implementation of different sorts of rights, such as the right to privacy.

These issues can be partly solved through insurance and reinsurance. There are already insurer companies that not only cover specific IoT risks considered at different operational levels, but more broadly the cascading effects generated by the cyber-physical nature of IoT infrastructure that can span across vast geographical areas and interact with the physical environment and various logical functions. In terms of pandemic management, there is also a value for medical business models. Some insurance companies cover the business aspects because they can distribute the risks across their portfolio in ways that allow a further decrease in average loss values. But given the speed of COVID-19 pandemic, it is unclear how fast can insurers adapt to the new risks. Many insurers simply back away with concerns of significant unpredicted and unmanageable losses.

**Why is the risk of connected devices difficult to assess?**

The risk from coupled and connected systems in pandemic management presents challenges in autonomous medical data collection, storing, processing and analysis. Here, we outline some of the inherent challenges in pandemic monitoring though coupled and connected systems.

- **Difficulties in assessing cyber risk:** The installation of new low-cost connected devices and sensors is, in some instances, not considered an IT or medical function. Connected devices and sensors serve a very diverse set of functions, ranging from simple automation such as room occupancy motion sensors for switching on the lights, to vastly more complex automation such as thermally regulating a building or ensuring its security. Installation of such sensors could be an operational task performed by the buildings and maintenance teams. Hence, in some instances, cyber risk from connected devices is invisible to cyber risk managers. Another example is retrofitting where IoT solutions are implemented on existing legacy systems. The justification for retrofitting is to reduce the cost of implementing new technologies. Also, since IoT evolves so fast, retrofitting is also justified to reduce the cost of new IoT technologies that are quickly becoming obsolete. But this creates security problems. Old legacy systems are based on older security programs, sometimes working with simple and often shared passwords and system accounts that are easy to breach. In this context, IoT technology, and digitalization in general, provides solutions and extend well beyond for the existing legacy systems to manufacturing floors and other production activities and consumers. The logical assumption is that connected devices dealing with biosensed or medical data for public health during a pandemic would face similar, if not greater, difficulties. Yet the potential benefits are immense; there are many examples from around the world on how existing medical information systems can be extended and enhanced through connected devices, including automatic diagnosis[1] for conditions unrelated to COVID-19; monitoring and supervision with live tracking systems[2]; or even virtual clinics[3]. One could argue that the real value of digitalization in medical systems depends largely on a strong correlation between cybersecurity and innovation. Global data governance is struggling to keep up with the fast evolution, especially in IoT cyber risks from non-traditional data, such as facial recognition data, facilities access data, and industrial control system data. However, such systems are already in place and operational in many countries. Currently, such systems are used for security reasons, and it is hard to see how it would be any different if the same system is used for pandemic management. The main reasonable concern seems to be the new information that could emerge from analyzing medical data collected for pandemic management, with deep learning and artificial intelligence algorithms.
- **Difficulties in assessing cyber risk from feeding medical data to deep learning and artificial intelligence algorithms:** The difficulties in identifying the risk from deep learning and artificial intelligence (AI) emerge from the limitations of such assessments on existing non-medical systems. The proposed digitalization in medical systems can also be seen as a dynamic automated predictive cognitive system supported by real-time intelligence for cyber medical analytics. This requires dynamic analytics of cyber-attack threat event frequencies to predict the cyber risk magnitudes of medical data loss, and/or alternative interpretations of the medical data. Despite the requirements for a predictive model built upon mathematical and statistical methods, there are currently no mathematical models that enable the quantitative assessment of cyber risk in any sector, including the medical sector. In our recent publication on this topic [5], we discovered that the lack of probabilistic data leads to qualitative cyber risk assessment approaches, where the outcome represents a speculative assumption [6]. Emerging quantitative models are effectively designed with ranges and confidence intervals based on expert opinions, and not probabilistic data [7].

---

[1] https://www.ncbi.nlm.nih.gov/pmc/articles/PMC6510889/
[2] https://coronaboard.kr/en/
[3] https://www.mobihealthnews.com/news/asia-pacific/ping-good-doctor-launches-commercial-operation-one-minute-clinics-china

However, quantitative risk impact estimation is needed for making decisions on topics such as estimating cybersecurity, cyber risk, and cyber insurance. Without a dynamic, real-time probabilistic risk data and cyber risk analytics enhanced with AI, these estimations can be outdated and imprecise. Hence, the impact of cyber risk on digital medical systems could be costly, and cybersecurity not necessarily effective. The value of cyber risk real-time data in non-medical systems can be explained in economic terms, where the level of cybersecurity is based on economic value. In medical systems, economic value is not the primary concern; instead, the focus is on the patient's safety and privacy. Therefore, even in times of pandemics, the AI integration in the communications network and the relevant cybersecurity technology must evolve in an ethical way that humans can understand, while maintaining the maximum trust and privacy of the users. The co-ordination of cyber protection and AI analytics of personal data from connected devices must be reliable to prevent abuse from insider threats, organized crime, terror organizations, or state-sponsored aggressors. Given the lack of assessment on data privacy in existing medical systems for pandemic management, we could drive a comparison from the private sector. Data risk has been encouraging the private sector to take steps to improve the management of confidential and proprietary information (i.e. customer or financial data), intellectual property, and PII (Personally Identifiable Information). Companies that are interested in obtaining new revenue streams from data have pursued innovative and cost-effective ways to protect such data. Therefore, the digitalization of medical systems needs to be designed similarly as the private sector has been in recent times. One additional level of security that needs to be anticipated that could evolve from such a new system is the analysis of the threat event frequency, with a dynamic and self-adapting AI. This would empower the design of a cognition engine mechanism for predicting the data loss magnitude through the control, analysis, distribution, and management of probabilistic data. While this would enhance the security of digital medical systems, it would also help future efforts by governments and the private sector to improve the management of confidential and proprietary information.

- **Cyber risk standards on data risk from connected devices:** Despite many efforts, there are no uniform standards governing data risk from connected devices. It is unlikely that such uniform standards can be developed on time for COVID-19 pandemic management. There are discussions on this topic and in the future, we can certainly expect such standards. However, this could be years ahead, while the current IoT operating model is based on shared responsibility. Given the rapid growth of the IoT shared ecosystem and the lack of guidance, businesses have already started building and applying their standards and protocols. The digitalization of medical systems for pandemic management could follow existing standards and protocols developed by the business community. The potential negative implications of such developments could hinder the value of autonomous data sharing by connected devices. One of the main strengths of IoT in global pandemic management is the ability to aggregate data from different sources and in different formats and connect to various networks using different protocols. The lack of common, unified, and global standards governing the IoT, creates significant barriers to the interoperability of connected devices in sharing medical data.
- **Security paradigms for connected devices in pandemic management:** Traditionally, cybersecurity has been separated into three paradigms, secure, vigilant, and resilient. The secure paradigm is focused on preventing certain risks from occurring. In terms of digital pandemic management, prevention must include multi-layered protection in the form of coupled systems. Coupled systems increase protection from invisible weaknesses, such as different interpretation of collected data. The vigilance paradigm refers to securing a digital pandemic management system in a method that can resist attacks over time. With the IoT technologies continuously evolving, it is not enough to simply have a security strategy digital pandemic management. Since the threats are changing, the strategy should also be changing. Thirdly, resiliency refers to how quickly the recovery process can enable normal operations of the system. This capability is essential for designing digital systems for pandemic management systems, in the context in which the first two paradigms, security, and vigilance, cannot guarantee that some form of failure would not occur.

- **Risk management of connected devices in pandemic management:** With connected devices speeding into the medical system during pandemic management, there is an urgent need for digital security officers that would oversee the increasingly connected critical infrastructure, data production processes, and smart data analytics. When considering the risk assessment approaches discussed so far, it becomes clear that the increase in medical data connectivity implies higher risks. Reducing open connections reduces cyber risk. But the IoT is a solution based on multiple connections. Although the open connections are anonymized, they still increase the attack surface. This creates a conflict between value and risk. However, there are methods to mitigate this situation, such as building security in the design process [8], known as Secure by Design principles[4], or considering how security is handled by the device across its lifetime and possible varying contexts of use, known as Security Ergonomics by Design principles [9].  Security considerations need to be applied early in the design development process. To secure IoT devices early in the design development process requires an increase in the cost of manufacturing (e.g. secure by default), production (e.g. secure by design), installation (e.g. secure by resilience), and maintenance (e.g. security updates). Such costs, unless applied through some form of unified and global standards, can hinder the competitiveness of the IoT in a digital medical system on a national level. Many non-medical industries require that expensive risk engineering assessment should be done throughout the entire lifecycle of the product. The digitalization of pandemic management needs to follow a similar approach. Since time is of the essence during pandemics, one solution would be to only use connected devices that are secure by default, design, resilience, and enabled for security updates. This would, however, limit the value of connected devices in pandemic management. Hence, the solutions could be found in the two-level system, where data from less secured devices is anonymized, and only used for a limited function.
- **The simple solution for data from less secured connected devices:** To safeguard from the 'invisible' IoT cyber risks, an integrated approach to cybersecurity is required in the digitalization of pandemic management. Decentralized security fails to assess how IoT connects operations in unexpected ways. The scale and scope of IoT data collected are often underestimated, including the risks from such data being assessed by third parties, which is often the case. In contrast, an integrated approach to security (e.g. ISA 3000) would assure that most IoT risks could be prevented before they even occur. Another relatively simple solution is to integrate operational capabilities with multi-layered cyber risk management. This could take the form of loosely coupled pandemic management systems. Such an approach can reduce the risk of widespread failure triggered by a single device. To prevent risk from retrofitting, with the rise of new technologies and new threats, the legacy systems will soon become incapable to be upgraded with the newest security that prevents new threats. It seems that the innovation process will resolve this category of cyber risk because it will simplify compliance and address specific operational technology need. In terms of resilience, a simple solution would be to create a fail-safe system. In such a system, malicious artificial intelligence though IoT devices could create a denial of service by themselves[5] and a failure in one element can disrupt the entire pandemic management system.

---

[4] https://assets.publishing.service.gov.uk/government/uploads/system/uploads/attachment_data/file/775559/Secure_by_Design_Report_.pdf

**Final remarks**

The COVID-19 pandemic has resulted in different countries developing digital surveillance approaches for pandemic management, some of which operating autonomously to collect, analyze, and share data. Currently, the digitalization of COVID-19 pandemic management is occurring at a fast rate across the globe, with the integration of automated and autonomous connected devices feeding real-time data to artificial intelligence algorithms. While the digitalization of medical systems presents strong value for pandemic management, efforts need to be focused on solutions that deliver value for pandemic management, and not on designing systems that expose personal data to cyber risk and could make things worst. There are safe and ethical ways of using artificial intelligence algorithms and data from connected devices for pandemic management. But the design of digital systems for pandemic management, need to anticipate that connected devices create new unpredictable and often invisible cyber risks that are currently unregulated and frequently ignored. Considering that these are new technologies that are evolving at a very fast rate, almost every new design a digital pandemic monitoring system can be classified as high risk. Many such digitalization approaches seem to be threading in uncertain waters (e.g. China), taking risks without fully understanding the impact, and operating with a rather hopeful strategy. Also, there is a lack of appropriate cyber insurance policies to transfer risk and the lack of standards and regulations to govern these new medical systems. Given these conditions, the separation of risks becomes urgent. With the appropriate separation of systems according to the potential risks, at least the medical professionals could keep parts of the pandemic management system operational during cyber-attacks, or during data privacy loss. As things stand at present, many digital pandemic management systems are driven by the monitoring opportunities and treatment potential. This leads to a scenario where developers of such systems can ignore threats, continue to chase opportunities and hope for the best, but that won't stop hackers exploit their opportunities, and interfering along the process.

---

[5] https://splinternews.com/this-guys-light-bulb-performed-a-dos-attack-on-his-enti-1793846000

**Keywords:** Digital pandemic management; COVID-19; cyber risk during pandemics; IoT connected systems.

**BIOS & Photo:**

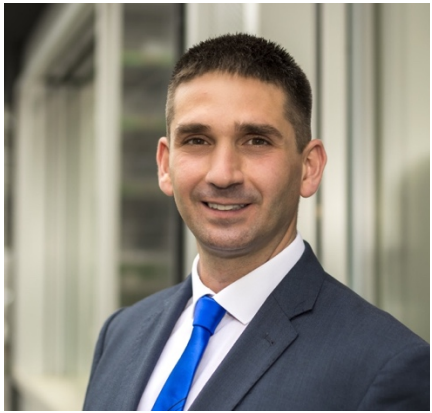

Petar Radanliev is a Post-Doctoral Research Associate at the University of Oxford. He obtained his Ph.D at University of Wales in 2014 and continued with Postdoctoral research at Imperial College London, Massachusetts Institute of Technology and the University of Oxford. His current research focusses on artificial intelligence, internet of things, cyber risk analytics and the value/impact of cyber risk.

Published in IEEE internet of things
News May 2020: https://iot.ieee.org/newsletter/may-2020/digitalization-of-covid-19-pandemic-management-and-cyber-risk-from-connected-systems

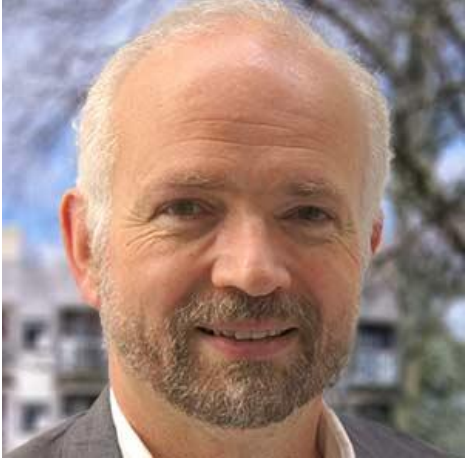

David De Roure is a Professor of e-Research at University of Oxford. He obtained his PhD at University of Southampton in 1990 and went on to hold the post of Professor of Computer Science, later directing the UK Digital Social Research programme. His current research focusses on social machines, Internet of Things and cybersecurity. He is a Fellow of the British Computer Society and the Institute of Mathematics and its Applications.

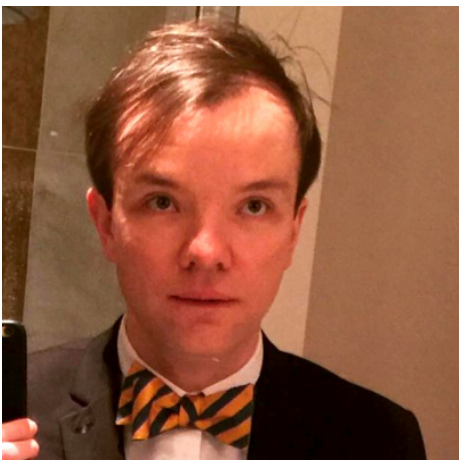

Max Van Kleek is Associate Professor of Human-Computer Interaction with the Department of Computer Science, at the University of Oxford. He works in the Software Engineering Programme, to deliver course material related to interaction design, the design of secure systems, and usability. His current project is designing new Web-architectures to help people re-gain control of information held about them "in the cloud", from fitness to medical records. He received his Ph.D. from MIT CSAIL in 2011.